\documentclass[superscriptaddress,twocolumn,amsmath,amssymb, prb]{revtex4}

\newcommand*{\ket}[1]{\left| #1 \right\rangle}
\newcommand*{\bra}[1]{\left\langle{#1}\right|}

\begin{document}
\title{Quantum mechanics of a free particle from properties of the Dirac delta function} 

\author{Denys I. Bondar}
\email{dbondar@sciborg.uwaterloo.ca}
\affiliation{University of Waterloo, Waterloo, Ontario N2L 3G1, Canada}

\author{Robert R. Lompay}
\email{rlompay@gmail.com}
\affiliation{Department of Theoretical Physics, Uzhgorod National University, Uzhgorod 88000, Ukraine}

\author{Wing-Ki Liu}
\email{wkliu@sciborg.uwaterloo.ca}
\affiliation{University of Waterloo, Waterloo, Ontario N2L 3G1, Canada}
\affiliation{Department of Physics, The Chinese University of Hong Kong,
Shatin, NT, Hong Kong}

\begin{abstract}
Based on the assumption that the probability density of finding a free particle is independent of position, we infer the form of the eigenfunction for the free particle, $\bra{x} p \rangle = \exp(ipx/\hbar)/\sqrt{2\pi\hbar}$. The canonical commutation relation between the momentum and position operators and the Ehrenfest theorem in the free particle case are derived solely from differentiation of the delta function and the form of $\bra{x} p \rangle$. 
\end{abstract}

\maketitle

The Dirac delta function\cite{Jackson2008} 
 is widely used in classical physics to describe the mass density of a point particle, the charge density of a point charge,\cite{Namias1977, Dennis1995, Aguirregabiria2002, Boykin2003} and the probability distribution of a random variable.\cite{Siegman1979, Gillespie1981, Gillespie1983} Quantum mechanical systems for which the potential is a delta function are, as a rule, exactly solvable.\cite{Atkinson1975, Foldy1976, Senn1988, Demkov1988, Goldstein1994, Albeverio2005, Gilbert2006}

The delta function is not a function in the usual sense. It is not even correct to define it as a limit of some ordinary functions (it can be represented as the ``weak'' limit of a sequence of functions). The delta function is a distribution, that is, a linear continuous functional defined on the space of ``good'' functions.\cite{Gelfand1964} Even though this definition might not be very appealing at first sight, it leads to consistent and fruitful mathematics.\cite{Gelfand1964} The theory of distributions allows us to perform linear operations on distributions as if they were ordinary functions. One result is the rule for differentiation of the delta function,\cite{Boykin2003} which we will show has important consequences such as the canonical commutation relation between the operators of momentum and position and the Ehrenfest theorems for a free particle. Our derivations use the relation for a free particle given by
\begin{equation}\label{bra_x_ket_p}
\bra{x} p \rangle = \exp(ipx/\hbar)/\sqrt{2\pi\hbar},
\end{equation}
where $\ket{x}$ and $\ket{p}$ are eigenstates of the position $\hat{x}$ and momentum $\hat{p}$ operators, respectfully (we consider only one dimension). 

We first present an intuitive derivation of Eq.~\eqref{bra_x_ket_p} based on the assumption that the probability density of a free particle is independent of its position. The purpose of our derivations is to demonstrate the utility and power of the theory of distributions, and to give a quantum mechanical interpretation of the mathematical properties of the delta function.

For simplicity, we list all the properties of the delta function that we will employ is this paper\cite{Schiff1968}
\begin{align}
x\delta'(x) & = - \delta(x), \label{RuleOfDifferentiation} \\
 \delta(-x) &= \delta(x), \label{EvenProperty} \\
 \int\!\frac{dx}{2\pi}\, x^n e^{-i\omega x} &= i^n \delta^{(n)} (\omega), \label{FourierRepresentation} \\
 x\delta(x-a) &= a\delta(x-a). \label{XDeltaXProperty}
\end{align}

We first present our ``derivation'' of Eq.~(\ref{bra_x_ket_p}), which depends on the properties of the delta function. We shall view the Dirac bra-ket notation as merely a convenient notation for eigenfunctions. The following two identities follow from the properties of eigenfunctions of self-adjoint operators
\begin{align}
 1 = \!\int\! dx \ket{x}\bra{x} &= \!\int\! dp \ket{p}\bra{p}, \label{Eigenfunc_ident1}\\
 \langle p \ket{p'} &= \delta (p-p'), \quad \langle x \ket{x'} = \delta(x-x'). \label{Eigenfunc_ident2}
\end{align}
The eigenfunction of a free particle is defined as the inner product $\langle x \ket{p}$. This quantity can be calculated either by postulating the explicit form of the operators $\hat{x}$ and $\hat{p}$ or by making another assumption. We shall select the second path. Experiments indicate that the probability density to find a free particle does not depend on its position. Thus,
\begin{equation}
\left| \langle x \ket{p} \right|^2 = \mbox{constant}.
\end{equation}
Hence, we conclude that
\begin{equation}\label{x_p_product_general}
\langle x \ket{p} = C \exp[i f(x,p)],
\end{equation}
where $C$ is a real constant and $f(x,p)$ is a smooth and real valued function. 
If we sandwich the right-hand side of Eq.~(\ref{Eigenfunc_ident1}) between $\langle x |$ and $|x' \rangle $ and use Eqs.~(\ref{Eigenfunc_ident2}) and (\ref{x_p_product_general}), we obtain
\begin{equation}\label{IntegralEq_for_f_general}
\delta (x-x') = \!\int\! dp \, \langle x \ket{p} \langle p \ket{x'} = C^2 \!\!\int\! dp \, e^{if(x,p) - if(x',p)},
\end{equation}
which can be considered to be the integral equation for the unknown function $f(x,p)$. For $x\neq x'$ we have
\begin{equation}
\label{IntegralEq_for_f_special}
\int\! dp \, e^{if(x,p) - if(x',p)} = 0 \qquad (x\neq x').
\end{equation}
Because $f(x,p)$ is sufficiently smooth, we can represent it as 
\begin{equation}\label{ExpansionF}
f(x,p) = g_1(x)p + g_2(x)p^2 + g_3(x)p^3 + \ldots
\end{equation}
If only the leading term is kept, we have
\begin{equation}
\int\! dp \, e^{ i[g_1(x) - g_1(x')]p} = 2\pi \delta( g_1(x) - g_1(x')),
\end{equation}
which satisfies Eq.~(\ref{IntegralEq_for_f_special}). If we truncate the expansion (\ref{ExpansionF}) after the $n$th term (for arbitrary $n \geq 2$), we obtain the integral
\begin{equation}
\int\! dp \, \exp\left( i\sum_{k=1}^n g_k p^k \right).
\end{equation}
This type of integral is known as a diffraction integral,\cite{Gilmore1981} and does not satisfy Eq.~(\ref{IntegralEq_for_f_special}) in general. Thus we assume 
$
f(x,p) = g_1(x)p
$.

We sandwich the middle expression of Eq.~(\ref{Eigenfunc_ident1}) between $\langle p' |$ and $|p \rangle $ and use Eqs.~(\ref{x_p_product_general}) and (\ref{Eigenfunc_ident2}) and the property (\ref{EvenProperty}) to find
\begin{equation}
\delta(p-p') = \!\int dx \, \langle p' \ket{x}\langle x \ket{p} = C^2 \!\!\int dx \, e^{if(x,p) - if(x,p')}.
\end{equation}
Instead of Eq.~(\ref{ExpansionF}), we expand $f(x,p)$ as a power series in $x$, and use the previous argument to conclude that $f(x,p)$ must be linear in $x$. Therefore, $f(x,p) = cpx$, where $c$ is a real constant. If we replace the left-hand side of Eq.~(\ref{IntegralEq_for_f_general}) by the Fourier representation of the delta function (\ref{FourierRepresentation}), we find
\begin{equation}\label{EqualityForCc}
\int\!\frac{dp}{2\pi\hbar} \, e^{ip(x-x')/\hbar} = C^2\!\! \int\! dp\, e^{icp(x-x')}.
\end{equation}
Therefore $C=1/\sqrt{2\pi\hbar}$ and $c=1/\hbar$, so that Eq.~(\ref{x_p_product_general}) reduces to Eq.~(\ref{bra_x_ket_p}). The constant $\hbar$  was introduced in Eq.~(\ref{EqualityForCc}) for dimensional purposes. 

Next we derive the canonical commutation relation for the position and momentum operators. We substitute $x \to x'-x$ into Eq.~(\ref{RuleOfDifferentiation}) and use Eq.~(\ref{EvenProperty}) to obtain
\begin{equation}\label{PreliminaryRelationForComutator}
(x' - x) \delta' (x'-x) = - \delta(x-x').
\end{equation}
Recall that
\begin{align}
\bra{x} \hat{p} \ket{x'} &= \!\int\! dp \, \bra{x}\hat{p}\ket{p}\bra{p}x' \rangle\\
&= \!\int\! \frac{dp}{2\pi\hbar} \, p e^{ip(x-x')/\hbar} = i\hbar\delta'(x'-x),
\end{align}
where Eq.~(\ref{FourierRepresentation}) was used in the last step. Thus,
\begin{align}
 (x' - x) \delta' (x'-x) &= x' \delta'(x' -x) - x \delta'(x'-x)\\
 &= \left( x' \bra{x}\hat{p}\ket{x'} -x \bra{x}\hat{p}\ket{x'} \right)/(i\hbar)\\
 &= \left( \bra{x}\hat{p}\hat{x}\ket{x'} - \bra{x}\hat{x}\hat{p}\ket{x'} \right)/(i\hbar)\\
 &= \bra{x} [\hat{p}, \hat{x}] \ket{x'} /(i\hbar).
\end{align}
We then employ Eq.~(\ref{PreliminaryRelationForComutator}) and the equality $\delta(x-x') = \bra{x} x' \rangle$ and obtain
\begin{equation}\label{XP_CommutationalRelation}
\bra{x} [\hat{p}, \hat{x}] \ket{x'} = \bra{x}(-i\hbar) \ket{x'},
\end{equation}
or 
\begin{equation}\label{XP_CommutationalRelation}
 [\hat{p}, \hat{x}] = -i\hbar,
\end{equation}
which is the commutation relation for the operators of position and momentum.

If we substitute $x \to p - p'$ into Eq.~(\ref{RuleOfDifferentiation}) and multiply both sides by $(p + p')/2$, we obtain 
\begin{equation}
\label{this1}
( p^2 - p'^2) \delta' (p-p')/2 = -(p+p') \delta(p-p')/2.
\end{equation}
Application of Eq.~(\ref{XDeltaXProperty}) to the right-hand side of Eq.~\eqref{this1} leads to
\begin{equation}
( p^2 - p'^2) \delta' (p-p')/2m = -p' \delta(p-p')/m,
\end{equation}
which can be written as
\begin{equation}
\frac{d}{dt} e^{it(p^2 -p'^2)/2m\hbar}\,i\hbar\delta'(p-p') = e^{it (p^2 -p'^2)/2m\hbar}\,p' \delta(p-p')/m.
\end{equation}
Because
\begin{equation}
\bra{p} \hat{x} \ket{p'} = \!\int \!\frac{dx}{2\pi\hbar}\, x e^{i(p'-p)x/\hbar} = i\hbar\delta' (p-p'),
\end{equation}
we obtain
\begin{equation}\label{EhrenfestTheoremForXandP}
m\frac{d}{dt} \bra{t, p} \hat{x} \ket{p', t} = \bra{t, p} \hat{p} \ket{p', t},
\end{equation}
where $\ket{p, t} = e^{-i p^2 t/(2m\hbar)} \ket{p}$ is the time-dependent eigenfunction of a free particle. Equation~(\ref{EhrenfestTheoremForXandP}) is the Ehrenfest theorem for a free particle. 

According to Eqs.~(\ref{RuleOfDifferentiation}) and (\ref{XDeltaXProperty}), $x^2 \delta'(x) = -x\delta(x) = 0$, and hence
\begin{equation}
(p - p')^2 \delta'(p-p') = 0.
\end{equation}
This relation is the starting point of the following equivalent transformations:
\begin{align}
i^2(p^2 - p'^2)^2 \delta' (p-p') /(2m\hbar)^2 &= 0 \\
\frac{d^2}{dt^2} e^{i t (p^2 - p'^2)/2m\hbar} \delta'(p - p') &= 0\\
\frac{d^2}{dt^2} e^{i t (p^2 - p'^2)/2m\hbar} \bra{p} \hat{x} \ket{p'} &= 0.
\end{align}
Therefore, we have obtained the Ehrenfest theorem for the ``acceleration'' for a free particle:
\begin{equation}\label{EhrenfestTheoremForA}
\frac{d^2}{dt^2} \bra{t, p} \hat{x} \ket{p', t} = 0.
\end{equation}
Equation (\ref{EhrenfestTheoremForA}) implies that there is no acceleration in this case.

Our derivation of the Ehrenfest theorems is not generalizable to a non-free particle case because the standard derivation of the Ehrenfest relations requires knowledge of the Schr\"{o}dinger equation.

In summary, we have demonstrated that the eigenfunction of a free particle can be obtained from the assumption of spacial uniformity of the probability density to find the particle. The canonical commutation relation between the operators of the momentum and position and the Ehrenfest theorems in the free particle case were derived from the rule for differentiation of the delta function and the eigenfunction of a free particle. We did not postulate the forms of the operators of the position and momentum or employ the Schr\"{o}dinger equation.

\end{document}